\documentclass[12pt]{article}
\newcommand{\be}{\begin{equation}}
\newcommand{\ee}{\end{equation}}
\newcommand{\ben}{\begin{equation*}}
\newcommand{\een}{\end{equation*}}
\newcommand{\ar}{\begin{array}}
\newcommand{\arn}{\end{array}}

\newcommand{\vk}{\vec{k}}
\newcommand{\vks}{\vec{k}^{\;2}}
\newcommand{\q}{\vec{q}}
\newcommand{\qs}{\vec{q}^{\;2}}
\newcommand{\qp}{\vec{q}^{\;\prime}}
\newcommand{\qps}{\vec{q}^{\;\prime\; 2}}
\newcommand{\x}{\vec{r}}
\newcommand{\xs}{\vec{r}^{\;2}}
\newcommand{\xp}{\vec{r}^{\;\prime}}

\def\pnot{\mbox{${\not{\hbox{\kern-3.0pt$p$}}}$}}
\def\qnot{\mbox{${\not{\hbox{\kern-2.0pt$q$}}}$}}
\def\enot{\mbox{${\not{\hbox{\kern-2.0pt$e$}}}$}}
\def\knot{\mbox{${\not{\hbox{\kern-2.0pt$k$}}}$}}

\def\fun#1#2{\lower3.6pt\vbox{\baselineskip0pt\lineskip.9pt\ialign
{$\mathsurround=0pt#1\hfil##\hfil$\crcr#2\crcr\sim\crcr}}}

\begin{document}

\begin{titlepage}

\begin{center}
{\bf On the coordinate representation of NLO BFKL$^{~\ast}$}
\end{center}
\vskip 0.5cm
\centerline{V.S.~Fadin$^{a\,\dag}$, R.~Fiore$^{b\,\ddag}$, A. Papa$^{b\,\dag\dag}$}
\vskip .6cm

\centerline{\sl $^{a}$ Budker Institute of Nuclear Physics, 630090 Novosibirsk,
Russia}
\centerline{\sl Novosibirsk State University, 630090 Novosibirsk, Russia}
\centerline{\sl $^{b}$ Dipartimento di Fisica, Universit\`a della Calabria,}
\centerline{\sl Istituto Nazionale di Fisica Nucleare, Gruppo
collegato di Cosenza,} \centerline{\sl Arcavacata di Rende, I-87036 Cosenza, Italy}
\vskip 2cm

\begin{abstract}
The ``non-Abelian'' part of the quark contribution to the BFKL
kernel in the next-to-leading order (NLO) is found in the
coordinate representation by direct transfer of the contribution
from the momentum representation where it was calculated before.
The results obtained are used for the examination of conformal
properties of the NLO BFKL kernel and of the relation between the
BFKL and color dipole approaches.

\end{abstract}
\vfill \hrule \vskip.3cm \noindent $^{\ast}${\it Work supported
in part by the Russian Fund of Basic Researches and in part by
Ministero Italiano dell'Istruzione, dell'Universit\`a e della
Ricerca.} \vfill $
\begin{array}{ll} ^{\dag}\mbox{{\it e-mail address:}} &
\mbox{FADIN@INP.NSK.SU}\\
^{\ddag}\mbox{{\it e-mail address:}} &
\mbox{FIORE@CS.INFN.IT}\\
^{\dag\dag}\mbox{{\it e-mail address:}} &
\mbox{PAPA@CS.INFN.IT}\\
\end{array}
$

\end{titlepage}

\vfill \eject

\section{Introduction}

The BFKL approach~\cite{BFKL}, based on the gluon Reggeization,
gives a common basis for the theoretical description of high
energy processes with fixed (not growing with energy) momentum
transfers. This approach  is well developed now in the
next-to-leading approximation (NLA). In particular, the kernel of
the BFKL equation is found  in the next-to-leading order (NLO) not
only for the forward scattering~\cite{FL98},  i.e. for $t=0$ and
the color singlet in the $t$--channel, but also for arbitrary
momentum transfer $t$ and any possible color state in the
$t$--channel~\cite{FFP99,FG00,FF05}. Particularly interesting is
the color singlet state, which is considered below, because physical
particles are colorless. All the  results mentioned above
are obtained in the momentum representation. There are at least
two reasons for considering the singlet BFKL kernel in the
coordinate representation in the transverse space.

First, just in this representation the BFKL equation   in the
leading approximation exhibits the famous property of conformal
invariance~\cite{Lipatov:1985uk}, which is extremely important for
finding solutions of the equation. Therefore conformal properties
of the NLO BFKL are very  interesting.  Evidently, the conformal
invariance is violated in the NLA by the renormalization; the
question is if the renormalization is the only source of
violation.

Second, it is the representation in which the color dipole
approach to high energy scattering~\cite{dipole}, very popular
now, is formulated. An advantage of the color dipole approach is a
clear physical picture of the high energy processes. Moreover,
this approach is naturally applied not only at low parton
densities, but also in the saturation regime~\cite{GLR83}, where
equations of evolution of parton densities with energy become
nonlinear. In general, there is an infinite hierarchy of coupled
equations~\cite{Balitsky,CGC}. In the simplest case, when the
target is a large nucleus,  it is reduced to the BK
(Balitsky-Kovchegov) equation~\cite{Balitsky}. A clear
understanding of the relation between these two approaches is very
important.  It could help in further development of the
theoretical description of small-$x$ processes. Unfortunately, the
``native'' representations for these two approaches are different:
for the color dipole approach it is the coordinate representation
in the transverse space, whereas for the BFKL approach it is the
momentum one. Therefore the relation between these two approaches
is not quite transparent, although  it is
affirmed~\cite{dipole,Balitsky}, that in the linear regime the
color dipole gives the same results as the BFKL approach for the
color singlet channel.

The leading order color singlet BFKL kernel has been investigated
in the coordinate representation in details~\cite{Lipatov:1985uk}
before the advent of the dipole approach. The relation between
BFKL and color dipole in the leading order was analyzed recently
in Ref.~\cite{Bartels:2004ef}. In this paper we extend this analysis to
the NLO. We plan to obtain both quark and gluon parts of the
kernel in the dipole approach by direct transformation of the BFKL
kernel in the momentum representation to the coordinate
representation. In this paper we consider the ``non-Abelian'' part
of the quark contribution to the NLO BFKL kernel.

\section{The leading order kernel in coordinate representation}

In this Section we consider the BFKL kernel at the leading order.
We will use the following notation: $\qp_i$ and $\q_i$, $i=1,2$,
represent the transverse momenta of Reggeons in initial and final
$t$-channel states, while $\xp_i$ and $\x_i$ are the corresponding
conjugate coordinates. The state normalization is \be
\label{normalization}\langle \q|\qp\rangle=\delta(\q-\qp)\;,
\;\;\;\;\; \langle \x|\xp\rangle=\delta(\x-\xp)\;, \ee so that \be
\langle\x|\q\rangle=\frac{e^{i\q\;\x}}{(2\pi)^{1+\epsilon}}\;, \ee
where $\epsilon=(D-4)/2$; $D-2$ is the dimension of the
transverse space and is taken different from $2$ for the
regularization of divergences. Note that  in our previous papers
we denoted the initial (final) momenta as $\q_1$ and $-\qp_1$
($\q_2$ and $-\qp_2$) and used the normalization
$\langle\q|\qp\rangle=\qs\delta(\q-\qp)$. We will use also the
notation $\q=\q_1+\q_2,\;\;\qp=\qp_1+\qp_2;
\;\;\vk=\q_1-\qp_1=\qp_2-\q_2$. The BFKL kernel in the operator
form is written as
\begin{equation}\label{operator of the BFKL kernel}
\hat{\cal K}=\hat{\omega}_1+\hat{\omega}_2+ \hat{\cal K}_r\;,
\end{equation}
where \be\label{trajectory ff} \langle \q_{i}|\hat{\omega}_i|
\qp_{i}\rangle=\delta(\q_i-\qp_i)\omega(-\qs_i )\;, \ee
with $\omega(t)$ the gluon Regge trajectory, and $\hat{\cal K}_r$
represents real particle production in Reggeon collisions. The
$s$-channel discontinuities of scattering amplitudes for the
processes $A+B\rightarrow A^\prime +B^\prime$ have the form
\be\label{discontinuity representation}
-4i(2\pi)^{D-2}\delta(\q_A-\q_B)\mbox{disc}_s{\cal A}_{AB}^{A'B'}=\langle
A^\prime \bar A|e^{Y\hat{\cal
K}}\frac{1}{\hat{\q}^{\;2}_1\hat{\q}^{\;2}_2}|\bar B^\prime
B\rangle\;. \ee In this equation $Y=\ln(s/s_0)$, $s_0$ is an appropriate
energy scale, $\;\;q_A=p_{A'}-p_A, \;\;q_B=p_B-p_{B'}$, and
\be\label{kernel ff} \langle \q_{1},\q_{2}|\hat{\cal K}|
\qp_{1},\qp_{2}\rangle=\delta(\q-\qp)\frac{1}{\qs_1\qs_2}{\cal
K}(\q_1,\qp_1;\q)\;, \ee \be\label{impact BB} \langle
\q_{1},\q_{2}|\bar B^\prime
B\rangle=4p_B^-\delta(\q_{B}-\q_{1}-\q_{2}){\Phi}_{B'B}(\q_{1},\q_{2})\;,
\ee \be\label{impact AA} \langle A^\prime \bar
A|\q_{1},\q_{2}\rangle=
4p_A^+\delta(\q_{A}-\q_{1}-\q_{2}){\Phi}_{A'A}(\q_{1},\q_{2})\;.
\ee where $p^{\pm}=(p_0\pm p_z)/\sqrt 2$;   the kernel ${\cal
K}(\q_1,\qp_1;\q)$ and the impact factors $\Phi$ are expressed
through the Reggeon vertices according to Ref.~\cite{FF98}. Note that
the appearance of the factors
$(\hat{\q}^{\;2}_1\hat{\q}^{\;2}_2)^{-1}$ in (\ref{discontinuity
representation}) and  $(\qs_1\qs_2)^{-1}$ in (\ref{kernel ff}) cannot
be explained by a change of the normalization~(\ref{normalization}).
At this point we have to recognize that
there is a substantial freedom in the definition of the kernel.
Indeed, the expression (\ref{discontinuity representation}) is not
changed by the transformation
\begin{equation}\label{kernel transformation}
\hat{\cal K}\rightarrow \hat{\cal O}^{-1}\hat{\cal K}\hat{\cal
O}~,\;\; \langle A^\prime \bar A|\rightarrow \langle A^\prime \bar
A|\hat{\cal O}~, \;\;\frac{1}{\hat{\q}^{\;2}_1\hat{\q}^{\;2}_2}|\bar B^\prime
B\rangle \rightarrow {\hat{\cal
O}^{-1}}\frac{1}{\hat{\q}^{\;2}_1\hat{\q}^{\;2}_2}|\bar B^\prime
B\rangle\;,
\end{equation}
with arbitrary nonsingular operator $\hat{\cal O}$. Actually we use a
kernel related with the one defined in Ref.~\cite{FF98} by such  transformation
with $\hat{\cal O}=(\hat{\q}^{\;2}_1\hat{\q}^{\;2}_2)^{1/2}$. The
reason is that  in the leading order the kernel which  is
conformal invariant and is simply related to the dipole kernel is
not the kernel defined in Ref.~\cite{FF98}, but just the kernel
 $\hat{\cal K}$ in (\ref{kernel ff})~\cite{Lipatov:1985uk,Bartels:2004ef}.
Remind that in this paper we consider
scattering of colorless objects and therefore only the color
singlet state in the $t$--channel.

In the leading order the trajectory in the momentum representation
is given by
\begin{equation}
\langle \q_i|\hat{\omega}_i|\qp_i\rangle=-\delta(\q_i-\qp_i)
\frac{g^2N_c\qs_i}{2(2\pi)^{D-1}} \int\frac{d^{D-2}r}{\vec
r^{\,2}(\vec q_i-\vec r)^{2}} =-\delta(\q_i-\qp_i)g^2 \frac{N_c
\Gamma(1-\epsilon)}{(4 \pi)^{D/2}}
\frac{\Gamma^2(\epsilon)}{\Gamma(2\epsilon)} (\vec
q_i^{\,2})^\epsilon  \label{LO trajectory}
\end{equation}
and the ``real'' part of the kernel by
\begin{equation}
\langle \q_1,\q_2|\hat{{\cal K}}_r|\qp_1,\qp_2\rangle =
\delta(\q-\qp)\frac{g^2N_c }{(2\pi
)^{D-1}}\frac{1}{\qs_1\qs_2}\left( \frac{\vec q_1^{\,2}\vec
q_2^{\,\prime\, 2}+\vec q_2^{\,2}\vec
q_1^{\,\prime\,2}}{\vks}-\vec q^{\,2}\right)\;. \label{real born
kernel in p space}
\end{equation}
Accordingly, in the coordinate representation we obtain
\begin{equation}
\langle \x_1,\x_2|\hat{\omega}_1+\hat{\omega}_2|\xp_1,\xp_2\rangle
=\frac{g^2N_c\Gamma^2(1+\epsilon)}{8\pi^{3+2\epsilon}}\left[\frac{\delta(\x_1-
\xp_1)}{(\x_2-\xp_2)^{2(1+2\epsilon)}}+\frac{\delta(\x_2-
\xp_2)}{(\x_1-\xp_1)^{2(1+2\epsilon)}}\right] \label{virtual
born kernel in x space}
\end{equation}
and
\[
\langle \x_1,\x_2|\hat{{\cal K}}_r|\xp_1,\xp_2\rangle
=\frac{g^2N_c \Gamma^2(1+\epsilon)}{4\pi
^{3+2\epsilon}}\int{d^{D-2}\rho}\frac{(\x_1-\vec\rho)}
{(\x_1-\vec\rho)^{2(1+\epsilon)}}
\frac{(\x_2-\vec\rho)}{(\x_2-\vec\rho)^{2(1+\epsilon)}}
\]
\begin{equation}
\times\left(\delta(\x_1-\xp_1)-\delta(\xp_1-\vec\rho)\right)
\left(\delta(\x_2-\xp_2)-\delta(\xp_2-\vec\rho)\right)\;.
\label{real born kernel in x space}
\end{equation}
Therefore the BFKL kernel can be written as
\[
\langle \x_1,\x_2|\hat{{\cal K}}|\xp_1,\xp_2\rangle =\langle
\x_1,\x_2|\hat{{\cal K}}_{d}|\xp_1,\xp_2\rangle
\]
\begin{equation}
-\frac{g^2N_c\Gamma^2(1+\epsilon)}{8\pi^{3+2\epsilon}}\left[\frac{\delta(\x_1-
\xp_1)}{(\x_1-\xp_2)^{2(1+2\epsilon)}}+\frac{\delta(\x_2-
\xp_2)}{(\x_2-\xp_1)^{2(1+2\epsilon)}}-2\frac{\delta(\xp_1-
\xp_2)(\x_1-\xp_1)(\x_2-\xp_2)}{(\x_1-\xp_1)^{2(1+\epsilon)}(\x_2-\xp_2)^{2
(1+\epsilon)}}\right]\;,  \label{singlet BFKL born kernel}
\end{equation}
where
\[
\langle \x_1,\x_2|\hat{{\cal K}}_{d}|\xp_1,\xp_2\rangle =\frac{g^2
N_c\Gamma^2(1+\epsilon)}{8\pi^{3+2\epsilon}}\int
d^{2+2\epsilon}\rho
\left(\frac{(\x_1-\vec\rho)}{(\x_1-\vec\rho)^{2(1+\epsilon)}}-
\frac{(\x_2-\vec\rho)}{(\x_2-\vec\rho)^{2(1+\epsilon)}}\right)^2
\]
\begin{equation}
\times\left(\delta(\x_1-\xp_1)\delta(\xp_2-\vec\rho)+\delta(\x_2-\xp_2)
\delta(\xp_1-\vec\rho)-\delta(\x_1-\xp_1)
\delta(\x_2-\xp_2)\right) \label{dipole born kernel}
\end{equation}
is just the dipole kernel in the $(D-2)$-dimensional space. It can
be seen from (\ref{singlet BFKL born kernel}) that the BFKL and
dipole kernels are not equivalent. However, if we consider
scattering of colorless objects, there is freedom in the
definition of the kernel~\cite{Lipatov:1985uk,Bartels:2004ef}. The
matter is that for colorless objects the impact factors $\langle
A^\prime \bar A|$ in (\ref{discontinuity representation}) have the
property of ``gauge invariance'': they turn into zero at zero
transverse momenta, i.e. $ \Phi_{A'A}(\vec 0, \q)=\Phi_{A'A}(\q,\vec
0)=0$. This means that $\langle A^\prime \bar A|\Psi\rangle=0$ for
any $|\Psi\rangle$, if $\langle \q_1,\q_2|\Psi\rangle$ contains
$\delta(\q_1)$ or $\delta(\q_2)$ (or, equivalently, in the
coordinate space that $\langle \x_1,\x_2|\Psi\rangle$ does not
depend either on $\x_1$ or on $\x_2$).  As it can be seen from
(\ref{operator of the BFKL kernel}), (\ref{LO trajectory}) and
(\ref{real born kernel in p space}), the BFKL kernel conserves
this property, i.e. if $\langle A^\prime \bar A|$ is ``gauge
invariant'', then also $\langle A^\prime \bar A|\hat{{\cal K}}$ is
``gauge invariant''. This property of the BFKL kernel is ensured
by the vanishing of its ``real'' part  $\langle
\q_1,\q_2|\hat{{\cal K}}_r|\qp_1,\qp_2\rangle$ (\ref{real born
kernel in p space}) at $\qp_1=0$ or $\qp_2=0$ and permits to
change in (\ref{discontinuity representation}) the input
\begin{equation}\label{input}
|\mbox{In}\rangle\equiv({\hat{\q}^{\;2}_1\hat{\q}^{\;2}_2})^{-1}|\bar
B^\prime B\rangle
\end{equation}
for $|\mbox{In}_d\rangle$, where $|\mbox{In}_d\rangle$ has the
``dipole'' property $\langle \x, \x|\mbox{In}_d\rangle=0$ (in the
momentum representation
\\$\int d^{D-2}q_1d^{D-2}q_2\langle \q_1,\q_2|\mbox{In}_d\rangle =0$).
Indeed, this can be done using substitutions of the type
\[
\langle \x_1,\x_2|\mbox{In}\rangle\rightarrow \langle \x_1,
\x_2|\mbox{In}\rangle -a\langle \x_1, \x_1|\mbox{In}\rangle
-(1-a)\langle \x_2, \x_2|\mbox{In}\rangle
\]
or, in the momentum representation,
\[
\langle \q_1,\q_2|\mbox{In}\rangle\rightarrow \langle
\q_1,\q_2|\mbox{In}\rangle-\left[a\delta(\q_2)+(1-a)\delta(\q_1)\right]\int
d^{D-2}q'_1d^{D-2}q'_2\langle \qp_1,\qp_2|\mbox{In}\rangle
\delta(\q_1+q_2-\qp_1-\qp_2)\;,
\]
which do not change the matrix element~(\ref{discontinuity
representation}). But for inputs of the ``dipole'' kind the last
three terms in the BFKL kernel (\ref{singlet BFKL born kernel})
can be omitted: the first two of them owing to the gauge
invariance of $\langle A^\prime \bar A|\hat{{\cal K}}^n$, and the
third because of $\hat{{\cal K}}_d$ conserves the ``dipole''
property, since  $\langle \x_1,\x_2|\hat{{\cal K}}_d
|\xp_1,\xp_2\rangle$ turns into zero at $\x_1=\x_2$ (see
(\ref{dipole born kernel})). After that the BFKL kernel turns to
the dipole one. Note, however, that it does not mean equivalence
of the kernels; in particular, their Green's functions are
different. Actually, the BFKL kernel acts in a wider space of
functions than the dipole one.

Thus, the BFKL and dipole kernels are not related by a simple
Fourier transform. However, by calculating  matrix elements of the
type (\ref{discontinuity representation})  with ``gauge
invariant'' impact factors $\langle A^\prime \bar A|$, one can
change the input (\ref{input}) into another one,
$|\mbox{In}_d\rangle$, with the ``dipole'' property $\langle \x,
\x|\mbox{In}_d\rangle=0$. Then, due to the fact that $ \langle
A^\prime \bar A|\hat{{\cal K}}^n$ is ``gauge invariant'' as well
as  $ \langle A^\prime \bar A|$, one can  add to $\langle
\x_1\x_2|\hat{{\cal K}}|\xp_1\xp_2\rangle $ terms not depending on
{$\x_1$} and {$\x_2$} (in the momentum space proportional  to
{$\delta(\q_1)$ or $\delta(\q_2)$}). They can be chosen in such a
way that after their addition the part of the kernel not
containing $\delta(\xp_1- \xp_2)$ becomes conserving the
``dipole'' property. After that the ``dipole'' property of the
$|\mbox{In}_d\rangle$ permits to omit the terms proportional  to
{$\delta(\xp_1- \xp_2)$} (in the momentum space depending not
separately on {$\qp_1$} and {$\qp_2$}, but only on the sum
{$\qp_1+\qp_2=\q_1+\q_2$}). In such a way we come from the BFKL to
the dipole kernel.

The use of the dipole kernel must not lead neither to infrared,
nor to ultraviolet singularities. The kernel $\hat{{\cal K}}_{d}$
satisfies this requirement. Indeed, for $\epsilon=0$ we have from
(\ref{dipole born kernel})
\begin{equation}
\langle \x_1,\x_2|\hat{{\cal K}}_{d}|\Psi\rangle =\frac{g^2
N_c}{8\pi^{3}}\int d^{2}\rho
\frac{(\x_1-\x_2)^2}{(\x_1-\vec\rho)^{2}(\x_2-\vec\rho)^{2}}
\left(\langle \x_1,\vec\rho|\Psi\rangle+\langle
\vec\rho,\x_2|\Psi\rangle-\langle \x_1,\x_2|\Psi\rangle\right)\;.
\label{dipole born kernel at D=4}
\end{equation}
The absence of infrared singularities is evident. Ultraviolet
singularities cancel taking into account the ``dipole'' property
$\langle \x,\x|\Psi\rangle=0$.

The  dipole form of the kernel is just the form which exhibits the
conformal invariance.  Conformal transformations in the
two-dimensional space $\x=(x,y)$ can be written as
\begin{equation}
z\rightarrow \frac{az+b}{cz+d}\;, \;\;\;
\end{equation}
where $z=x+iy, \;\; a,b,c,d$ are complex numbers, with $ad-bc\neq
0$. Under these transformations, one has
\[
z_1-z_2\rightarrow \frac{z_1-z_2}{(cz_1+d)(cz_2+d)}(ad-bc)\;,
\]
\begin{equation}
dzdz^*\rightarrow
dzdz^*\frac{|ad-bc|^2}{\left|(cz+d)^2\right|^2}\;,
\end{equation}
so that the conformal invariance of the  dipole kernel is evident
from (\ref{dipole born kernel at D=4}).

In the momentum representation the kernels are connected by the
relation
\begin{equation}\label{K and K d in p} \langle \q_1\q_2|\hat{{\cal
K}}|\qp_1\qp_2\rangle =\langle \q_1\q_2|\hat{{\cal
K}}_{d}|\qp_1\qp_2\rangle
-\delta(\q-\qp)\left[\delta(\q_2)\omega(\qp_2)+\delta(\q_1)\omega(\qp_1)
+\frac{g^2N_c}{(2\pi)^{3+2\epsilon}}\frac{2\q_1\q_2}{\qs_1\qs_2}\right]\;.
\end{equation}
This equation can be obtained by the direct transformation of
(\ref{singlet BFKL born kernel}) in the momentum representation.
From (\ref{K and K d in p}) and from (\ref{operator of the BFKL
kernel}), (\ref{LO trajectory}) and (\ref{real born kernel in p
space}) it follows  that
\begin{equation}
\int d^{D-2}q_1d^{D-2}q_2\langle\q_1\q_2|\hat{\cal
K}_{d}|\qp_1\qp_2\rangle =0\;,
\end{equation}
which is the expression of the ``dipole'' property of $\hat{\cal
K}_{d}$ in the momentum space. This means that the dipole kernel
does not satisfy to the ``bootstrap
condition''~\cite{Bartels:2004ef}, which can be written as
\begin{equation}
\int d^{D-2}q_1d^{D-2}q_2\langle\q_1\q_2| (\hat{\cal K}
+\hat{\omega}_1+\hat{\omega}_2- 2\omega(t)
 )|\qp_1\qp_2\rangle=0\;.
\end{equation}

\section{The unrenormalized NLO kernel in coordinate representation}

In this Section, and from now on, we consider only the quark contribution
to the NLO kernel. Moreover,
we use the large $N_c$ limit, where only the ``non-Abelian'' part
of the ``real'' contribution survives, so that the contribution is
strongly simplified~\cite{FFP99}.

The quark contribution $ \omega^Q$ to the trajectory appears at
the two-loop level only. For the case of $n_f$ massless quark
flavours  we have~\cite{FFP99}
\[
\langle
\q|\hat{\omega}^Q|\qp\rangle=\delta(\q-\qp)\frac{8g^4N_cn_f\Gamma^2
\left( 1-\epsilon\right) \Gamma ^2\left(2+\epsilon\right)\Gamma
^2\left(1+\epsilon\right)}{(4\pi )^{4+2\epsilon}\Gamma \left(
4+2\epsilon\right)\Gamma \left(1+2\epsilon\right) } (\vec
q^{\,2})^{2\epsilon}\frac{1}{\epsilon^2}
\]
\begin{equation}
\times\left(1-\frac{3\Gamma(1-2\epsilon)\Gamma^2(1+2\epsilon)}
{2\Gamma^2(1-\epsilon)\Gamma(1+\epsilon)\Gamma(1+3\epsilon)}\right)
\;. \label{Q part of trajectory}
\end{equation}
In the coordinate representation we get
\begin{equation}
\langle
\x_1\x_2|\hat{\omega}^Q_1+\hat{\omega}^Q_2|\xp_1\xp_2\rangle
 =-a_v
\left[\frac{\delta(\x_1-
 \xp_1)}{\epsilon(\x_2-\xp_2)^{2(1+3\epsilon)}}+\frac{\delta(\x_2-
 \xp_2)}{\epsilon(\x_1-\xp_1)^{2(1+3\epsilon)}}\right]\;,
\label{virtual Q part  in x space}
\end{equation}
where
\begin{equation}\label{a v and kappa}
a_v=2a_r\left(\kappa-\frac{3}{2}\right), \;\;
\kappa=\frac{\Gamma^2(1-\epsilon)\Gamma(1+\epsilon
)\Gamma(1+3\epsilon )}{\Gamma(1-2\epsilon )\Gamma^{2}(1+2\epsilon
)}\;,
\end{equation}
\begin{equation}\label{a r}
a_r=\frac{2g^{4}N_c n_{f}2^{1+4\epsilon}\Gamma(1+\epsilon
)\Gamma(1+2\epsilon )}{(4\pi )^{2+\epsilon
}(2\pi)^{D-1}}\frac{\Gamma^{2}(2+\epsilon )}{\Gamma (4+2\epsilon
)}~.
\end{equation}
The contribution of real $q\bar q$ production is given by Eq.~(47)
of Ref.~\cite{FFP99}~\footnote{In Eq.~(47) of Ref.~\cite{FFP99}
there is actually a misprint in one sign, easily detectable by
comparison with Eq.~(48) of that paper or with Eq.~(\ref{real  Q
part in p space}) of this work, which are correct.}:
\[
\langle \q_1\q_2|\hat{\cal K}^{Q}_r|\qp_1\qp_2\rangle =
\delta(\q-\qp)\frac{2g^{4}N_c n_{f}}{(4\pi )^{2+\epsilon
}}\frac{\Gamma (1-\epsilon )}{\epsilon(2\pi )^{D-1}
}\frac{\Gamma^{2}(2+\epsilon )}{\Gamma (4+2\epsilon
)\vec{q}_{1}^{\:2}\vec{q}_{2}^{\:2}}
\]
\[
\times \Biggl\{
2{\vec{k}}^{2(\epsilon-1)}(\vec{q}_{1}^{\:2}\vec{q}_{2}^{\:\prime
\:2}+\vec{q}_{2}^{\:2}\vec{q} _{1}^{\:\prime \:2})+\vec q^{\:2}
\left(2\vec q^{\:2\epsilon}- \vec{q}_{1}^{\:2\epsilon}
-\vec{q}_{1}^{\:\prime\:2\epsilon}-\vec{q}_{2}^{\:2\epsilon}
-\vec{q}_{2}^{\:\prime\:2\epsilon} \right) \Biggr.
\]
\begin{equation}
\Biggl.  -\frac{(\vec{q}_{1}^{\:2}\vec{q}_{2}^{\:\prime
\:2}-\vec{q}_{2}^{\:2} \vec{q}_{1}^{\:\prime
\:2})}{{\vec{k}}^{2}}\left(
\vec{q}_{1}^{\:2\epsilon}-\vec{q}_{1}^{\:\prime\:2\epsilon}
-\vec{q}_{2}^{\:2\epsilon}+\vec{q}_{2}^{\:\prime\:2\epsilon
}\right) \Biggr\} \;. \label{real  Q part in p space}
\end{equation}
It is easy to see that the expression in the curly brackets
vanishes when any of the $\vec{q}_{i}$'s or $\vec{q}_{i}^{\:\prime
}$'s tends to zero. Vanishing at $\vec{q}_{i}^{\:\prime }=0$ is
important for us because it means that $\hat{\cal K}^{Q}$
conserves the ``gauge invariance'' of the impact factor $\langle
A^\prime \bar A|$ as well as the  kernel in the leading order.

In the coordinate representation $\hat{\cal K}^{Q}_r$ can be
presented as (for details, see Ref.~\cite{FFP})
\[
\langle \x_1\x_2|\hat{{\cal K}}^{Q}_r|\xp_1\xp_2\rangle = a_r
\Biggl[\delta(\x_1-\xp_1)\delta(\x_2-\xp_2)\int d^{D-2}\rho
f(\x_1,\x_2;\rho)+\delta(\xp_1-\xp_2) f_0(\x_1,\x_2;\xp_1)
\]
\begin{equation}\label{real kernel through f}
+\delta(\x_1-\xp_1)
f_1^r(\x_1,\x_2;\xp_2)+\delta(\x_2-\xp_2)f_2^r(\x_1,\x_2;\xp_1)+
\frac{\Gamma(1+\epsilon)}{\pi^{1+\epsilon}}f(\x_1,\x_2;\xp_1,\xp_2)\Biggr]\;,
\end{equation}
where
\begin{equation}\label{function f}
 f(\x_1,\x_2;\vec\rho)=\frac{(\x_1-\vec\rho)(\x_2-\vec\rho)}
 {\epsilon(\x_1-\vec\rho)^{2(1+\epsilon)}(\x_2-\vec\rho)^{2(1+\epsilon)}
}\left(
\frac{1}{(\x_1-\vec\rho)^{2\epsilon}}+\frac{1}{(\x_2-\vec\rho)^{2\epsilon}}
\right)\;,
\end{equation}
\[
f_0(\x_1,\x_2;\vec\rho)= \frac{(\x_1-\vec\rho)(\x_2-\vec\rho)}
{\epsilon(\x_1-\vec\rho)^{2(1+\epsilon)}(\x_2-\vec\rho)^{2(1+\epsilon)}}
\left(\frac{1}{(\x_1-\vec\rho)^{2\epsilon}}+
\frac{1}{(\x_2-\vec\rho)^{2\epsilon}}\right)
\]
\[
+\frac{1}{(\x_1-\vec\rho)^{2(1+2\epsilon)}}
\left(\frac{1}{\epsilon(\x_2-\vec\rho)^{2\epsilon}}
-\frac{1}{\epsilon(\x_1-\x_2)^{2\epsilon}}\right)
+\frac{1}{(\x_2-\vec\rho)^{2(1+2\epsilon)}}
\left(\frac{1}{\epsilon(\x_1-\vec\rho)^{2\epsilon}}-
\frac{1}{\epsilon(\x_1-\x_2)^{2\epsilon}}\right)
\]
\begin{equation}\label{function f 0}
+2\int
\frac{d^{D-2}\x_3}{\pi^{1+\epsilon}}\frac{\Gamma(1+\epsilon)}
{(\x_3-\vec\rho)^{2(1+2\epsilon)}}
\frac{(\x_1-\x_3)(\x_2-\x_3)}{(\x_1-\x_3)^{2(1+\epsilon)}
(\x_2-\x_3)^{2(1+\epsilon)}}\;,
\end{equation}
\[
f^r_1(\x_1,\x_2;\vec\rho)=
-2\frac{(\x_1-\vec\rho)(\x_2-\vec\rho)}{
\epsilon(\x_1-\vec\rho)^{2(1+2\epsilon)}(\x_2-\vec\rho)^{2(1+\epsilon)}}
-\frac{1}{\epsilon(\x_1-\vec\rho)^{2(1+2\epsilon)}}
\left(\frac{1}{(\x_2-\vec\rho)^{2\epsilon}}
-\frac{1}{(\x_1-\x_2)^{2\epsilon}}\right)
\]
\begin{equation}\label{function f 1 r}
-\frac{(\x_1-\vec \rho)(\x_2-\vec\rho)}{\epsilon(\x_1-\vec
\rho)^{2(1+\epsilon)}(\x_2-\vec\rho) ^{2(1+2\epsilon)}}-\int
\frac{d^{D-2}\x_3}{\pi^{1+\epsilon}}
\frac{\Gamma(1+\epsilon)}{(\x_3-\vec\rho)^{2(1+2\epsilon)}}
\frac{(\x_1-\x_3)(\x_2-\x_3)}{(\x_1-\x_3)^{2(1+\epsilon)}
(\x_2-\x_3)^{2(1+\epsilon)}}\;,
\end{equation}
\begin{equation}\label{function f 2 r}
f^r_2(\x_1,\x_2;\vec\rho)=f^r_1(\x_2,\x_1;\vec\rho)\;,
\end{equation}
\begin{equation}\label{function f large }
f(\x_1,\x_2;\xp_1,\xp_2)=f(\x_1,\x_2)+1\leftrightarrow 2\;,
\end{equation}
\[
f(\x_1,\x_2)= \frac{1} {(\x_1-\xp_1)^{2(1+2\epsilon)}}
\frac{(\x_2-\xp_2)(\xp_2-\xp_1)}{(\x_2-\xp_2)^{2(1+\epsilon)}
(\xp_2-\xp_1)^{2(1+\epsilon)}}
\]
\begin{equation}\label{function f little}
+\left(\frac{1}{(\x_1-\xp_1)^{2(1+2\epsilon)}}
-\frac{1}{(\xp_1-\xp_2)^{2(1+2\epsilon)}} \right)
\frac{(\x_1-\xp_2)(\x_2-\xp_2)}{(\x_1-\xp_2)^{2(1+\epsilon)}
(\x_2-\xp_2)^{2(1+\epsilon)}}\;,
\end{equation}
where $1\leftrightarrow 2$ means $\x_1\leftrightarrow \x_2,
\;\;\xp_1\leftrightarrow \xp_2$. Note, that actually only the
integral
\begin{equation}\label{integral of f}
\int d^{D-2}\rho f(\x_1,\x_2;\rho)=
\frac{\pi^{1+\epsilon}}{\epsilon\Gamma(1+\epsilon)(\x_1-\x_2)^{4\epsilon}}
\end{equation}
is fixed, so that there is a large arbitrariness  in definition of
$f(\x_1,\x_2;\rho)$.

Now the total kernel is presented in the form (\ref{real kernel
through f}) with the substitution
$f_i^r(\x_1,\x_2;\vec\rho)\rightarrow f_i(\x_1,\x_2;\vec\rho)$,
where
\begin{equation}\label{function f i through f i_r}
f_i(\x_1,\x_2;\vec\rho)=f_i^r(\x_1,\x_2;\vec\rho)-\frac{a_v}{a_r}
\frac{1}{\epsilon(\x_j-\vec\rho)^{2(1+3\epsilon)}}\;,
\end{equation}
with $i=1,2$ and $j\neq i$.

Let us transform the part of the kernel without the term with
$\delta( \xp_1- \xp_2)$ to the ``dipole'' form using the freedom,
discussed above, to redefine it in matrix elements
(\ref{discontinuity representation}) with ``gauge invariant''
impact factors $\langle A^\prime \bar A|$. As well as in the
leading order we change the input (\ref{input}) for another one,
$|\mbox{In}_d\rangle$, with the ``dipole'' property $\langle \x,
\x|\mbox{In}_d\rangle=0$. Then, due to the fact that $\hat{\cal
K}^{Q}$ conserves the ``gauge invariance'' of the impact factor
$\langle A^\prime \bar A|$, we add to $\langle \x_1\x_2|\hat{{\cal
K}}|\xp_1\xp_2\rangle$ terms  not depending on {$\x_1$} and
{$\x_2$}. Moreover, we change $f(\x_1,\x_2;\vec\rho)$ using the
famous property of $(D-2)$--dimensional integrals
\begin{equation}\label{property of dimensional regularization}
\int \frac{d^{D-2}\rho}{(\vec\rho -\x_i)^\alpha}=0\;.
\end{equation}
As a result the considered part must have neither infrared, nor
ultraviolet singularities and possess the ``dipole'' property.

First we note that the functions $f_i(\x_1,\x_2;\vec\rho)$,
$i=1,2, $ have ultraviolet singularities at $\vec \rho =\x_j \;
j\neq i$ (the singularities at $\vec \rho =\x_i$ in separate terms
cancel each other) which can be removed by adding to
$f(\x_1,\x_2;\vec\rho)$ the term
\begin{equation}\label{addition to f}
\frac{a_v}{a_r}\left(\frac{1}{\epsilon(\x_1-\vec\rho)^{2(1+3\epsilon)}}+
\frac{1}{\epsilon(\x_2-\vec\rho)^{2(1+3\epsilon)}}\right)\;.
\end{equation}
After that the remaining singularities are infrared. Note that the
addition of the term (\ref{addition to f}) improves also the
infrared behavior of $f(\x_1,\x_2;\vec\rho)$. Since for
$(\vec\rho-\x_1)^2\simeq(\vec\rho-\x_2)^2 \gg (\x_1-\x_2)^2$ we
have
\begin{equation}\label{approx}
\int \frac{d^{D-2}\x_3}{\pi^{1+\epsilon}}\frac{\Gamma(1+\epsilon)}
{(\x_3-\vec\rho)^{2(1+2\epsilon)}}
\frac{(\x_1-\x_3)(\x_2-\x_3)}{(\x_1-\x_3)^{2(1+\epsilon)}
(\x_2-\x_3)^{2(1+\epsilon)}}
\end{equation}
\[
\simeq\frac{1}{\epsilon(\vec\rho-\x_1)^{2(1+2\epsilon)}(\x_1-\x_2)^{2\epsilon}}
-\frac{2\kappa}{\epsilon(\vec\rho-\x_1)^{2(1+3\epsilon)}}\;,
\]
where $\kappa$ is defined in (\ref{a v and kappa}),  the infrared
divergences in $f_i(\x_1,\x_2;\vec\rho)$ can be removed by adding
$1/[\epsilon(\x_i-\vec\rho)^{2(1+3\epsilon)}]$. Note that this
addition does not create ultraviolet singularities due to the
assumed ``dipole'' property of the input. The important fact is
that such addition provides the ``dipole'' property of
$f_i(\x_1,\x_2;\vec\rho)$, namely $f_i(\x,\x;\vec\rho)=0$, as it
follows immediately using~(\ref{approx}).

Let us turn now to $f(\x_1, \x_2; \xp_1,\xp_2)$. As it can be
seen, it has only infrared singularities.  We can remove them (not
creating new ones) performing the following transformation
\begin{equation}\label{transformation}
f(\x_1, \x_2)\rightarrow \tilde f(\x_1, \x_2)=f(\x_1,
\x_2)-af(\x_1, \x_1)-(1-a)f(\x_2, \x_2)\;.
\end{equation}
The new function $\tilde f(\x_1, \x_2)$ has the important
``dipole'' property $\tilde f(\x, \x)=0$, independently from the
value of $a$. In our case
\begin{equation}\label{f(r,r)}
f(\x, \x)=\frac{1} {(\x-\xp_1)^{2(1+2\epsilon)}}
\frac{(\x-\xp_2)(\xp_2-\xp_1)}{(\x-\xp_2)^{2(1+\epsilon)}
(\xp_2-\xp_1)^{2(1+\epsilon)}}
\]
\[
+ \frac{1}{(\x-\xp_2)^{2(1+2\epsilon)}}
\Biggl[\frac{1}{(\x-\xp_1)^{2(1+2\epsilon)}}
-\frac{1}{(\xp_1-\xp_2)^{2(1+2\epsilon)}}\Biggr]\;,
\end{equation}
so that $f(\x_1, \x_1)$ and $ f(\x_2, \x_2)$ are related by the
substitution $1\leftrightarrow 2$ (that means $\x_1\leftrightarrow
\x_2$ and  $\xp_1\leftrightarrow \xp_2$). As a consequence,
$f(\x_1,\x_2;\xp_1,\xp_2)$, see (\ref{function f large }), does
not depend on $a$, so that we can take any $a$. Thus we come to
the  representation (\ref{real kernel through f}), where the term
with $\delta(\xp_1-\xp_2)$ is omitted and in the other terms we
make the replacement $f\rightarrow \tilde f$,
\[
\tilde f(\x_1,\x_2;\vec\rho)=\frac{(\x_1-\vec\rho)(\x_2-\vec\rho)}
{\epsilon(\x_2-\vec\rho)^{2(1+\epsilon)}
(\x_1-\vec\rho)^{2(1+\epsilon)}}\left(
\frac{1}{(\x_1-\vec\rho)^{2\epsilon}}+\frac{1}{(\x_2-\vec\rho)^{2\epsilon}}
\right)
\]
\begin{equation}\label{function  tilde f 0}
+\frac{a_v}{a_r}\Biggl(\frac{1}{\epsilon(\x_1-\vec\rho)^{2(1+3\epsilon)}}
+\frac{1}{\epsilon(\x_2-\vec\rho)^{2(1+3\epsilon)}}\Biggr)\;,
\end{equation}
\[
\tilde f_1(\x_1,\x_2;\vec\rho)=
-2\frac{(\x_1-\vec\rho)(\x_2-\vec\rho)}{\epsilon
(\x_1-\vec\rho)^{2(1+2\epsilon)}(\x_2-\vec\rho)^{2(1+\epsilon)}}-
\frac{a_v}{a_r}\frac{1}{\epsilon(\x_2-\vec\rho)^{2(1+3\epsilon)}}+
\frac{1}{\epsilon(\x_1-\vec\rho)^{2(1+3\epsilon)}}
\]
\[
-\frac{1}{\epsilon(\x_1-\vec\rho)^{2(1+2\epsilon)}}
\left(\frac{1}{(\x_2-\vec\rho)
^{2\epsilon}}-\frac{1}{(\x_1-\x_2)^{2\epsilon}}\right)
-\frac{(\x_1-\vec \rho)(\x_2-\vec\rho)}{\epsilon(\x_1-\vec
\rho)^{2(1+\epsilon)}(\x_2-\vec\rho) ^{2(1+2\epsilon)}}
\]
\begin{equation}\label{function tilde f 1}
-\int \frac{d^{D-2}\x_3}{\pi^{1+\epsilon}}
\frac{\Gamma(1+\epsilon)}{(\x_3-\vec\rho)^{2(1+2\epsilon)}}
\frac{(\x_1-\x_3)(\x_2-\x_3)}{(\x_1-\x_3)^{2(1+\epsilon)}
(\x_2-\x_3)^{2(1+\epsilon)}}\;,
\end{equation}
\begin{equation}\label{function tilde f 2}
\tilde f_2(\x_1,\x_2;\vec\rho)=\tilde f_1(\x_2,\x_1;\vec\rho)\;,
\end{equation}
\[
\tilde f(\x_1,\x_2;\xp_1,\xp_2)=\tilde
f(\x_1,\x_2)+1\leftrightarrow 2\;,
\]
\begin{equation}\label{function f tilde}
\tilde f(\x_1,\x_2)= f(\x_1,\x_2)-f(\x_1,\x_1)\;,
\end{equation}
where $f(\x_1,\x_2)$ is defined in (\ref{function f little}).

\section{Renormalization of the NLO kernel in coordinate representation}

In order to get the renormalized quark contribution to NLO kernel,
we have to express the bare coupling $g$ in terms of the
renormalized one $g_\mu$ and to add the part proportional to $n_f$
coming from the coupling renormalization in the leading order
kernel (\ref{dipole born kernel}). In the $\overline{\mbox{MS}}$
scheme we have
\begin{equation}
g^2=g^2_\mu \mu ^{-2\mbox{\normalsize $\epsilon$}}\left[ 1+\left(
\frac{11}3-\frac 23\frac{n_f}N_c\right) \frac{\bar g_\mu
^2}{\epsilon }\right] \;,  \label{g and g mu}
\end{equation}
where
\begin{equation}
\bar g_\mu ^2=\frac{g_\mu ^2N_c\Gamma (1-\epsilon )}{(4\pi
)^{2+{\epsilon }}}\;. \label{bar g mu}
\end{equation}
Therefore the contribution coming from the  coupling
renormalization in the leading order kernel is obtained from
(\ref{dipole born kernel}) by the substitution
\begin{equation}\label{substitution for born}
\frac{g^2 N_c\Gamma^2(1+\epsilon)}{8\pi^{3+2\epsilon}}\rightarrow
-\frac{\bar g^4_\mu
n_f}{N_c}\frac{2^{2+2\epsilon}\mu^{-2\epsilon}}{\pi^{1+\epsilon}}
\frac{\Gamma^2(1+\epsilon)}{3\epsilon\Gamma(1-\epsilon)}=-\lambda
\frac{a_r}{\epsilon} \;,
\end{equation}
where
\begin{equation}\label{lambda}
\lambda=\frac{\Gamma(1-\epsilon)\Gamma(1+\epsilon)\Gamma(4+2\epsilon)}
{6 \cdot
2^{2\epsilon}\mu^{-2\epsilon}\Gamma(1+2\epsilon)\Gamma^2(2+\epsilon)}\;,
\end{equation}
and $a_r$ is expressed in terms of $\bar g_\mu$,
\begin{equation}\label{renormalized a R}
a_r=\frac{8\bar g_\mu^4\mu^{-4\epsilon}
2^{4\epsilon}\Gamma^{2}(2+\epsilon )\Gamma(1+\epsilon
)\Gamma(1+2\epsilon )n_{f}}{\pi^{1+\epsilon }\Gamma
(4+2\epsilon)\Gamma^2(1-\epsilon) N_c}\;.
\end{equation}
Then the quark contribution to the renormalized kernel in the
``dipole'' form is written as
\[
\langle \x_1\x_2|\hat{{\cal K}}^{Q}_d|\xp_1\xp_2\rangle = a_r
\Biggl[\delta(\x_1-\xp_1)\delta(\x_2-\xp_2)\int d^{D-2}\rho \bar
f(\x_1,\x_2;\rho)
\]
\begin{equation}\label{kernel through bar f}
+\delta(\x_1-\xp_1) \bar f_1(\x_1,\x_2;\xp_2)+\delta(\x_2-\xp_2)
\bar f_2(\x_1,\x_2;\xp_1)+
\frac{\Gamma(1+\epsilon)}{\pi^{1+\epsilon}} \bar
f(\x_1,\x_2;\xp_1,\xp_2)\Biggr]\;,
\end{equation}
where
\[
 \bar f(\x_1,\x_2;\vec\rho)=\frac{(\x_1-\vec\rho)(\x_2-\vec\rho)}
 {\epsilon(\x_2-\vec\rho)^{2(1+\epsilon)}
(\x_1-\vec\rho)^{2(1+\epsilon)}}\left(
\frac{1}{(\x_1-\vec\rho)^{2\epsilon}}+\frac{1}{(\x_2-\vec\rho)^{2\epsilon}}-2\lambda\right)
\]
\begin{equation}\label{function  bar f 0}
+\left(\frac{a_v}{a_r(\x_1-\vec\rho)^{2\epsilon}}+\lambda\right)
\frac{1}{\epsilon(\x_1-\vec\rho)^{2(1+2\epsilon)}}
+\left(\frac{a_v}{a_r(\x_2-\vec\rho)^{2\epsilon}}+\lambda\right)\frac{1}
{\epsilon(\x_2-\vec\rho)^{2(1+2\epsilon)}}\Biggr)\;,
\end{equation}
\[
\bar f_1(\x_1,\x_2;\vec\rho)=
-2\left(\frac{1}{(\x_1-\vec\rho)^{2\epsilon}}-\lambda\right)\frac{(\x_1-\vec\rho)(\x_2-\vec\rho)}{\epsilon
(\x_1-\vec\rho)^{2(1+\epsilon)}(\x_2-\vec\rho)^{2(1+\epsilon)}}
\]
\[
-
\left(\frac{a_v}{a_r(\x_2-\vec\rho)^{2\epsilon}}+\lambda\right)\frac{1}
{\epsilon(\x_2-\vec\rho)^{2(1+2\epsilon)}}+
\left(\frac{1}{(\x_1-\vec\rho)^{2\epsilon}}-\lambda\right)\frac{1}{\epsilon(\x_1
-\vec\rho)^{2(1+2\epsilon)}}
\]
\[
-\frac{1}{\epsilon(\x_1-\vec\rho)^{2(1+2\epsilon)}}\left(\frac{1}{(\x_2-\vec\rho)
^{2\epsilon}}-\frac{1}{(\x_1-\x_2)^{2\epsilon}}\right)
-\frac{(\x_1-\vec \rho)(\x_2-\vec\rho)}{\epsilon(\x_1-\vec
\rho)^{2(1+\epsilon)}(\x_2-\vec\rho) ^{2(1+2\epsilon)}}
\]
\begin{equation}\label{function bar f 1}
-\int
\frac{d^{D-2}\x_3}{\pi^{1+\epsilon}}\frac{\Gamma(1+\epsilon)}{(\x_3-\vec\rho)^{2(1+2\epsilon)}}
\frac{(\x_1-\x_3)(\x_2-\x_3)}{(\x_1-\x_3)^{2(1+\epsilon)}
(\x_2-\x_3)^{2(1+\epsilon)}}\;,
\end{equation}
\begin{equation}\label{function bar f 2}
\bar f_2(\x_1,\x_2;\vec\rho)=\bar f_1(\x_2,\x_1;\vec\rho)\;,
\end{equation}
\[
\bar f(\x_1,\x_2;\xp_1,\xp_2)=\tilde
f(\x_1,\x_2)+1\leftrightarrow 2\;,
\]
\[
\tilde f(\x_1,\x_2)= f(\x_1,\x_2)-f(\x_1,\x_1)\;,
\]
and $f(\x_1,\x_2)$ is defined in (\ref{function f little}). These
equations give the quark contribution to the dipole kernel at
arbitrary $D$.

The limit $\epsilon\rightarrow 0$ can be  easily taken.   We can
put $\epsilon=0$ in the common coefficient $a_r$ and in the ratio
$a_r/a_v$ (see (\ref{renormalized a R}), (\ref{bar g mu}) and
(\ref{a v and kappa})), getting
\begin{equation}\label{limit of a r}
a_r=\frac{\alpha_s^2(\mu)N_cn_f}{12\pi^3}\;,\;\;\;\;
\frac{a_r}{a_v}=-1\;,
\end{equation}
and then we can expand the integrand in $\epsilon$. Using
\[
\int \frac{d^{D-2}\rho}{\pi^{1+\epsilon}}\frac{\Gamma(1+\epsilon)}
{(\xp_1-\vec\rho)^{2(1+2\epsilon)}}
\frac{(\x_1-\vec\rho)(\x_2-\vec\rho)}{(\x_1-\vec\rho)^{2(1+\epsilon)}
(\x_2-\vec\rho)^{2(1+\epsilon)}}
\]
\begin{equation}
\simeq\frac{(\x_1-\xp_1)(\x_2-\xp_1)}{(\x_1-\xp_1)^2(\x_2-\xp_1)^2}
\left(-\frac{1}{\epsilon}+
\ln\left(\frac{(\x_1-\xp_1)^4(\x_2-\xp_1)^4}{(\x_1-\x_2)^2}\right)\right)\;,
\end{equation}
and
\begin{equation}\label{lambda at small epsilon}
\lambda\simeq
1+\epsilon\left(\frac{5}{3}-2\psi(1)+\ln\left(\frac{\mu^2}{4}\right)\right)\equiv
1-\epsilon\ln \xs_\mu~; \;\; \ln \xs_\mu=
-\frac{5}{3}+2\psi(1)-\ln\left(\frac{\mu^2}{4}\right)\;,
\end{equation}
we obtain finally
\[
 \bar f(\x_1,\x_2;\vec\rho)=\frac{1}{2}\Biggl[-\frac{(\x_1-\x_2)^2}
 {(\x_1-\vec\rho)^{2}(\x_2-\vec\rho)^{2}}\ln\left(
\frac{\x_\mu^{\:4}}{(\x_1-\vec\rho)^{2}(\x_2-\vec\rho)^{2}}\right)
\]
\begin{equation}\label{limit of function bar f 0}
+\frac{(\x_2-\vec\rho)^{2}-(\x_1-\vec\rho)^{2}}{(\x_1-\vec\rho)^{2}(\x_2-\vec\rho)^{2}}
\ln\left(
\frac{(\x_1-\vec\rho)^{2}}{(\x_2-\vec\rho)^{2}}\Biggr]\right)\;,
\end{equation}
\[
\bar f_1(\x_1,\x_2;\vec\rho)=
\frac{1}{2}\Biggl[\frac{(\x_1-\x_2)^2}
 {(\x_1-\vec\rho)^{2}(\x_2-\vec\rho)^{2}}\ln\left(
\frac{\x_\mu^{\:4}}{(\x_1-\vec\rho)^{2}(\x_1-\x_2)^{2}}\right)
\]
\begin{equation}\label{limit of function bar f 1}
+\frac{(\x_1-\vec\rho)^{2}-(\x_2-\vec\rho)^{2}}{(\x_1-\vec\rho)^{2}(\x_2-\vec\rho)^{2}}
\ln\left(
\frac{(\x_1-\vec\rho)^{2}(\x_1-\x_2)^{2}}{(\x_2-\vec\rho)^{4}}\right)\Biggr]\;,
\end{equation}
\begin{equation}\label{limit of function bar f 2}
\bar f_2(\x_1,\x_2;\vec\rho)=\bar f_1(\x_2,\x_1;\vec\rho)\;,
\end{equation}
\[
\bar f(\x_1,\x_2;\xp_1,\xp_2) =\frac{(\x_1-\x_2)(\xp_1-\xp_2)}
{(\x_1-\xp_1)^2(\x_2-\xp_2)^2(\xp_1-\xp_2)^2}+\frac{(\x_1-\x_2)^2}
{2(\x_1-\xp_2)^2(\x_2-\xp_2)^2}\Biggl(\frac{1}{(\xp_1-\xp_2)^2}
-\frac{1}{(\x_1-\xp_1)^2}\Biggr)
\]
\begin{equation}\label{limit of function bar f}
+\frac{(\x_1-\x_2)^2}
{2(\x_2-\xp_1)^2(\x_1-\xp_1)^2}\Biggl(\frac{1}{(\xp_1-\xp_2)^2}
-\frac{1}{(\x_2-\xp_2)^2}\Biggr).
\end{equation}
We notice that the conformal invariance is violated not only by
the renormalization.  We see also that the result of
transformation to the coordinate representation of the BFKL kernel
defined in the momentum representation by~(\ref{operator of
the BFKL kernel}), (\ref{trajectory ff}) and (\ref{kernel ff})
does not coincide with the result obtained recently
in Ref.~\cite{Balitsky:2006wa} by direct calculation of the quark
contribution to the dipole kernel in the coordinate
representation. This is evident from the presence of the function $\bar
f(\x_1,\x_2;\xp_1,\xp_2)$, which is absent at large $N_c$ in the
result of Ref.~\cite{Balitsky:2006wa}.  However, at this point we have to
remind about the freedom in the definition of the BFKL  kernel,
which was discussed at the beginning of Section~2. We have fixed
the operator $\hat{\cal O}$ in (\ref{kernel transformation}), but
transformations with $\hat{\cal O}=1-\hat O$, where $\hat O\sim
g^2$,  are still possible. At the NLO after such transformation we get
\begin{equation}\label{transformation at NLO}
\langle \x_1\x_2|\hat{{\cal K}}|\xp_1\xp_2\rangle \rightarrow
\langle \x_1\x_2|\hat{{\cal K}}|\xp_1\xp_2\rangle-\langle
\x_1\x_2|[\hat{{\cal K}}^{(B)},\hat O]|\xp_1\xp_2\rangle~,
\end{equation}
where $\hat{{\cal K}}^{(B)}$ is the leading order kernel.

If we take
\begin{equation}\label{operator o}
\hat O
=\frac{2g^2n_f}{(4\pi)^{2+\epsilon}}\frac{\Gamma(1-\epsilon)\Gamma^2(2+\epsilon)}
{\epsilon\Gamma(4+2\epsilon)}\left(\hat{\q}_1^{\;2\epsilon}+\hat{\q}_2^{\;2\epsilon}
\right)~,
\end{equation}
then, as a result of the transformation, all the functions $f$ in
(\ref{real kernel through f}) change to $f\rightarrow f+h$ ,
where
\begin{equation}\label{function h}
 h(\x_1,\x_2;\vec\rho)=0\;,
\end{equation}
\begin{equation}\label{function h 0}
h_0(\x_1,\x_2;\vec\rho)= \frac{(\x_1-\vec\rho)(\x_2-\vec\rho)}
 {\epsilon(\x_1-\vec\rho)^{2(1+\epsilon)}(\x_2-\vec\rho)^{2(1+\epsilon)}
}\left(
\frac{1}{(\x_1-\vec\rho)^{2\epsilon}}+\frac{1}{(\x_2-\vec\rho)^{2\epsilon}}
\right) \;,
\end{equation}
\[
h^r_1(\x_1,\x_2;\vec\rho)= -\frac{(\x_1-\vec\rho)(\x_2-\vec\rho)}{
\epsilon(\x_1-\vec\rho)^{2(1+\epsilon)}(\x_2-\vec\rho)^{2(1+2\epsilon)}}
-\frac{1}{\epsilon(\x_2-\vec\rho)^{2(1+2\epsilon)}}
\left(\frac{1}{(\x_1-\vec\rho)^{2\epsilon}}
-\frac{1}{(\x_1-\x_2)^{2\epsilon}}\right)
\]
\begin{equation}\label{function h 1}
-\int \frac{d^{D-2}\x_3}{\pi^{1+\epsilon}}
\frac{\Gamma(1+\epsilon)}{(\x_3-\vec\rho)^{2(1+2\epsilon)}}
\frac{(\x_1-\x_3)(\x_2-\x_3)}{(\x_1-\x_3)^{2(1+\epsilon)}
(\x_2-\x_3)^{2(1+\epsilon)}}\;,
\end{equation}
\begin{equation}\label{function h 2}
h^r_2(\x_1,\x_2;\vec\rho)=h^r_1(\x_2,\x_1;\vec\rho)\;,
\end{equation}
\begin{equation}\label{function h large }
h(\x_1,\x_2;\xp_1,\xp_2)=-f(\x_1,\x_2;\xp_1,\xp_2)\;.
\end{equation}
Note that the functions $h_i^r$ have neither ultraviolet nor
infrared singularities and possess the dipole property. Taking
into account that in the limit $\epsilon\rightarrow 0$ we have
\begin{equation}\label{limit of function h 1}
h^r_1(\x_1,\x_2;\vec\rho)=\frac
12\frac{(\x_1-\x_2)^{2}+(\x_1-\vec\rho)^{2}-(\x_2-\vec\rho)^{2}}
{(\x_1-\vec\rho)^{2}(\x_2-\vec\rho)^{2}} \ln\left(
\frac{(\x_1-\vec\rho)^{2}}{(\x_1-\x_2)^{2}}\right)\;,
\end{equation}
\begin{equation}\label{limit of function h 2}
h^r_2(\x_1,\x_2;\vec\rho)=h^r_1(\x_2,\x_1;\vec\rho)\;,
\end{equation}
we come to the conclusion that the transformation (\ref{transformation
at NLO}) with $\hat O$ defined in (\ref{operator o}) leads to the
substitutions in (\ref{limit of function bar f 0})-(\ref{limit of
function bar f})
\begin{equation}\label{f ---fb}
\bar f(\x_1,\x_2;\xp_1,\xp_2)\rightarrow 0~, \;\; \bar
f_{1,2}(\x_1,\x_2;\vec\rho)\rightarrow f_{b}(\x_1,\x_2;\vec\rho)~,
\end{equation}
where
\begin{equation}\label{limit of function f b 1 2}
f_{b}(\x_1,\x_2;\vec\rho)= \Biggl[\frac{(\x_1-\x_2)^2}
 {(\x_1-\vec\rho)^{2}(\x_2-\vec\rho)^{2}}\ln\left(
\frac{\x_\mu^{\:2}}{(\x_1-\x_2)^{2}}\right)
+\frac{(\x_1-\vec\rho)^{2}-(\x_2-\vec\rho)^{2}}{(\x_1-\vec\rho)^{2}(\x_2-\vec\rho)^{2}}
\ln\left(
\frac{(\x_1-\vec\rho)^{2}}{(\x_2-\vec\rho)^{2}}\right)\Biggr]\;.
\end{equation}
Up to the definition of the renormalization scale $\xs_\mu$ (which
also can be considered as the result of an appropriate
transformation) this function coincides with the one which appears in the
result of Ref.~\cite{Balitsky:2006wa}. As for $\bar
f(\x_2,\x_1;\vec\rho)$, we remind that only the integral
(\ref{integral of f}) is really defined. Without change of the
integral we can add to $f(\x_2,\x_1;\vec\rho)$ in (\ref{function
f}) the function $ h(\x_2,\x_1;\vec\rho)$
\[
 h(\x_1,\x_2;\vec\rho)=\frac{(\x_1-\vec\rho)(\x_2-\vec\rho)}
 {\epsilon(\x_1-\vec\rho)^{2(1+\epsilon)}(\x_2-\vec\rho)^{2(1+\epsilon)}
}\left(\frac{2}{(\x_1-\x_2)^{2\epsilon}}
-\frac{1}{(\x_1-\vec\rho)^{2\epsilon}}-\frac{1}{(\x_2-\vec\rho)^{2\epsilon}}
\right)
\]
\begin{equation}
-\frac{1}
 {\epsilon(\x_2-\vec\rho)^{2(1+2\epsilon)}
}\left(\frac{1}{(\x_1-\x_2)^{2\epsilon}}
-\frac{1}{(\x_1-\vec\rho)^{2\epsilon}} \right)-\frac{1}
 {\epsilon(\x_1-\vec\rho)^{2(1+2\epsilon)}
}\left(\frac{1}{(\x_1-\x_2)^{2\epsilon}}
-\frac{1}{(\x_2-\vec\rho)^{2\epsilon}} \right)\;,
\end{equation}
with the properties
\[
\int d^{2+2\epsilon}\rho \;h(\x_1,\x_2;\vec\rho)=0~,
\]
\[
h(\x_1,\x_2;\vec\rho)|_{\epsilon\rightarrow
0}=\frac{1}{2}\Biggl[\frac{(\x_1-\x_2)^2}
 {(\x_1-\vec\rho)^{2}(\x_2-\vec\rho)^{2}}\ln\left(
\frac{(\x_1-\x_2)^{4}}{(\x_1-\vec\rho)^{2}(\x_2-\vec\rho)^{2}}\right)
\]
\begin{equation}
+\frac{(\x_2-\vec\rho)^{2}-(\x_1-\vec\rho)^{2}}{(\x_1-\vec\rho)^{2}(\x_2-\vec\rho)^{2}}
\ln\left(
\frac{(\x_1-\vec\rho)^{2}}{(\x_2-\vec\rho)^{2}}\right)\Biggr]\;.
\end{equation}
As a result we have in~(\ref{limit of function bar f 0}) that $\bar
f(\x_1,\x_2;\vec\rho)\rightarrow - f_{b}(\x_1,\x_2;\vec\rho)$. After
this the dipole form of the kernel (\ref{transformation at NLO}),
(\ref{operator o}) coincides with the result of Ref.~\cite{Balitsky:2006wa}
(up to the definition of the renormalization scale).

\section{Direct transformation of the renormalized NLO kernel at D=4}

Till now, in order to be as general and rigorous as possible, we worked
starting with the BFKL kernel in the momentum representation at
arbitrary space-time dimension $D$. However, the results of
transformation to the coordinate representation in the physical
space-time dimension $D=4$ can be obtained in a much easier way,
if we start from the renormalized BFKL kernel at $D=4$ in a
specific form. To obtain this form let us use the renormalized
quark contribution to the BFKL kernel in the form~\cite{FFP99},
\[
\langle \q_1\q_2|\hat{\cal K}^Q_r|\qp_1\qp_2\rangle_{renorm}
=\delta(\q-\qp)F_r(\q_1, \qp_1; \q)~, \;\;
\]
\[
F_r(\q_1, \qp_1; \q)= \frac{\bar{g}_{\mu }^{4}\mu ^{-2\epsilon
}}{\pi ^{1+\epsilon }\Gamma (1-\epsilon
)}\frac{2n_{f}}{3N_c}\left\{ \frac{2}{\epsilon }\left(
\frac{6[\Gamma (2+\epsilon )]^{2}}{ \Gamma (4+2\epsilon )}\left(
{\frac{{\vec{k}}^{\;2}}{\mu ^{2}}}\right) ^{\epsilon }-{1}\right)
 \left(\frac{\vec{q}_{2}^{\:\prime \:2}}{\qs_2{\vec{k}
}^{2}}+\frac{\vec{q}_{1}^{\:\prime
\:2}}{\qs_1\vks}-\frac{\vec{q}^{\:2}}{\qs_1\qs_2}\right) \right.
\]
\begin{equation}\label{F r}
\left. + \frac{\vec{q}^{\:2}}{\qs_1\qs_2}\ln \left(
\frac{\vec{q}^{\:4}\vk^{ 4}}
{\vec{q}_{1}^{\:2}\vec{q}_{2}^{\:2}\qps_1\qps_2}\right)
-\left(\frac{\vec{q}_{2}^{\:\prime \:2}}{\qs_2{\vec{k}
}^{2}}-\frac{\vec{q}_{1}^{\:\prime \:2}}{\qs_1\vks}\right)\ln
\left(
\frac{\vec{q}_{1}^{\:2}\qps_2}{\vec{q}_{2}^{\:2}\qps_1}\right)
\right\}~.
\end{equation}
We also use the integral representation for the quark part of the
trajectory (taking into account the renormalization), which also can
be found in Ref.~\cite{FFP99},
\begin{equation}\label{omega as integral}
\omega^Q(-\qs_i)|_{renorm}=\int d^{2+2\epsilon} k F_\omega(\vk,
\q_i)~,
\end{equation}
where
\begin{equation}\label{F omega}
F_\omega(\vk, \q_i)=\frac{\bar g^{4}_\mu \mu^{-2\epsilon}\; }{\pi
^{1+\epsilon} \epsilon \Gamma (1-\epsilon )}\frac{2n_{f}}{3N_c}
\frac{\qs_i}{\vks(\q_i-\vk)^2}\left[1+\frac{6\Gamma^{2}(2+\epsilon
)}{\Gamma (4+2\epsilon )}
\left(\left(\frac{\qs_i}{\mu^{2}}\right)^\epsilon-\!\!\!
\left(\frac{\vks}{\mu^{2}}\right)^\epsilon-\!\!
\left(\frac{(\q_i-\vk)^2}{\mu^{2}}\right)^\epsilon\right)\right].
\end{equation}
Now, introducing a small cut-off $\lambda$ and making it tending to zero
after taking the limit $\epsilon\rightarrow 0$, we can write
$F_r=F_r(\theta(\lambda^2-\vks)+\theta(\vks-\lambda^2))$. Then in
the second region we can take the limit $\epsilon=0$ in (\ref{F
r}), whereas the contribution of the first region exactly cancel
the pieces of the trajectories coming from the integration regions
$\vks\leq \lambda^2$ and $(\q_i-\vk)^2\leq \lambda^2$ in
(\ref{omega as integral}). Outside these regions we can take the
limit $\epsilon=0$ also in $F_\omega(\vk, \q_i)$. Thus we come to
the kernel at $D=4$, where
\[
F_r(\q_1, \qp_1; \q)=
\frac{\alpha^2_s(\mu)}{16\pi^3}\frac{2 N_c n_{f}}{3}\left\{
{2}\left( \ln\left(\frac{{\vec{k}}^{\;2}}{\mu ^{2}}\right) -\frac
53\right)
 \left(\frac{\vec{q}_{2}^{\:\prime \:2}}{\qs_2{\vec{k}
}^{2}}+\frac{\vec{q}_{1}^{\:\prime
\:2}}{\qs_1\vks}-\frac{\vec{q}^{\:2}}{\qs_1\qs_2}\right) \right.
\]
\begin{equation}\label{F r D=4}
\left. + \frac{\vec{q}^{\:2}}{\qs_1\qs_2}\ln \left(
\frac{\vec{q}^{\:4}\vk^{ 4}}
{\vec{q}_{1}^{\:2}\vec{q}_{2}^{\:2}\qps_1\qps_2}\right)
-\left(\frac{\vec{q}_{2}^{\:\prime \:2}}{\qs_2{\vec{k}
}^{2}}-\frac{\vec{q}_{1}^{\:\prime \:2}}{\qs_1\vks}\right)\ln
\left(
\frac{\vec{q}_{1}^{\:2}\qps_2}{\vec{q}_{2}^{\:2}\qps_1}\right)
\right\}~,
\end{equation}
\begin{equation}\label{F omega D=4}
F_\omega(\vk, \q_i)=-\frac{\alpha^2_s(\mu)}{16\pi^3}
\frac{2 N_c n_{f}}{3}\frac{\qs_i}{\vks(\q_i-\vk)^2}\left(
\ln\left(\frac{{\vk^2(\q_i-\vk)^2}}{\mu ^{2}\qs_i}\right) -\frac
53\right)~.
\end{equation}
Of course, at that the virtual and real parts contain infrared
singularities. We have to remember that the singularities must be
regularized  by limitations on integration regions discussed above
or in an equivalent way.

Note that in the result of the transformation (\ref{transformation
at NLO}), (\ref{operator o})
\[
F_r(\q_1, \qp_1; \q)\rightarrow
\frac{\alpha^2_s(\mu)}{16\pi^3}\frac{4 N_c n_{f}}{3}\left\{
 \frac{\vec{q}_{2}^{\:\prime \:2}}{\qs_2{\vec{k}
}^{2}}\left(\ln\left(\frac{{\vec{k}}^{\;2}\qs_2}{\mu
^{2}\vec{q}_{2}^{\:\prime \:2}}\right)-\frac 53\right)
+\frac{\vec{q}_{1}^{\:\prime \:2}}{\qs_1\vks}
\left(\ln\left(\frac{{\vec{k}}^{\;2}\qs_1}{\mu
^{2}\vec{q}_{1}^{\:\prime \:2}}\right)-\frac 53\right) \right.
\]
\begin{equation}\label{F transformed  r D=4}
\left. - \frac{\vec{q}^{\:2}}{\qs_1\qs_2}\left(\ln \left( \frac
{\vec{q}_{1}^{\:2}\vec{q}_{2}^{\:2}}{\vec{q}^{\:2}\mu^{2}}\right)-\frac
53 \right)\right\}~,
\end{equation}
whereas the trajectory remains unchanged. Omitting the terms with
$\delta(\xp_1-\xp_2)$ we can write the transformed kernel
$\tilde{\cal K}^Q$ in the coordinate representation as
\[
\langle \x_1\x_2|\tilde{{\cal K}}^{Q}_d|\xp_1\xp_2\rangle =
\frac{\alpha^2_s(\mu)N_cn_f}{12\pi^3}
\Biggl[\delta(\x_1-\xp_1)\delta(\x_2-\xp_2)\int d^{2}\rho\,
g_0(\x_1,\x_2;\rho)
\]
\begin{equation}\label{tilde kernel through g}
+\delta(\x_1-\xp_1)  g(\x_1,\x_2;\xp_2)+\delta(\x_2-\xp_2)
g(\x_2,\x_1;\xp_1)\Biggr]\;.
\end{equation}
Let us define
\[
g(\x_1,\x_2)=\int\frac{d^2\rho}{(2\pi)}\frac{d^2k}{(2\pi)}\frac{d^2q}{(2\pi)}
\frac{d^2q^{\prime}}{(2\pi)}\Biggr[\frac{1}{\qs}\ln\frac{\vks}{\qps}
+\frac{1}{\vks}\ln\frac{\qs}{\qps}+\frac{2\vk\q}{\vks\qs}
\left(\ln\left(\frac{\vks\qs}{\qps\mu^2}\right)-\frac 53\right)
\Biggr]
\]
\begin{equation}
\times
e^{i\vk(\x_1-\vec\rho)+i\q(\x_2-\vec\rho)-i\qp(\xp_2-\vec\rho)}\;.
\end{equation}
Then the function  $g(\x_1,\x_2;\xp_2)$ can be written as
\begin{equation}\label{function  g}
g(\x_1,\x_2;\xp_2)=g(\x_1,\x_2)-\frac 12 g(\x_2,\x_2)-\frac 12
g(\x_1,\x_1).
\end{equation}
Here the first term comes from the real part, the second from the
trajectory, and the third is added ``by hand'', because its
contribution is zero due to the gauge invariance. It is not a
complicated task  to calculate $g(\x_1,\x_2)$. The only integral
which is not trivial is
\begin{equation}\label{integral}
\int\frac{d^2k}{(2\pi)}\frac{d^2q}{(2\pi)}\frac{\vk\q}{\vks\qs}
\ln(\vk+\q)^2e^{i\vk\x_1+i\q\x_2}
=\frac{\x_1\x_2}{\xs_1\xs_2}\left(-2\psi(1)-2\ln 2
+\ln\left(\frac{\xs_1\xs_2}{(\x_1-\x_2)^2}\right)\right)\;.
\end{equation}
As a result we have
\[
g(\x_1,\x_2)=-2\frac{(\x_1-\xp_2)(\x_2-\xp_2)}{(\x_1-\xp_2)^2(\x_2-\xp_2)^2}
\ln\left(\frac{\x_\mu^{\:2}}{(\x_1-\x_2)^2}\right)
\]
\begin{equation}\label{result for g 12}
+\frac{1}{(\x_1-\xp_2)^2}\ln\left(\frac{(\x_2-\xp_2)^2}{(\x_1-\x_2)^2}\right)
+\frac{1}{(\x_2-\xp_2)^2}\ln\left(\frac{(\x_1-\xp_2)^2}{(\x_1-\x_2)^2}\right)
\end{equation}
and $g(\x_1,\x_2;\xp_2)=f_b(\x_1,\x_2;\xp_2)$, with $f_b$ given in
(\ref{limit of function f b 1 2}).

To find the function $g_0$ is an even simpler task. Using the
representation
\[
\int d^2 k\frac{2}{\vks}\left(\ln\frac{\vks}{\mu^2}-\frac
53\right)e^{i\vk(\x_1-\x_2)}=-2\int d^2\rho
\frac{d^2k}{(2\pi)}\frac{d^2q}{(2\pi)}
\frac{\vk\q}{\vks\qs}\left(\frac 12 \ln\frac{\vks\qs}{\mu^4}-\frac
53\right)e^{i\vk(\x_1-\vec\rho)+i\q(\x_2-\vec\rho)}
\]
we have for the real part contribution $g_0(\x_1,\x_2)$ to
$g_0(\x_1,\x_2; \rho)$:
\begin{equation}
g_0(\x_1,\x_2)=
\frac{(\x_1-\vec\rho)(\x_2-\vec\rho)}{(\x_1-\vec\rho)^2(\x_2-\vec\rho)^2}
\ln\left(\frac{\x_\mu^{\:4}}{(\x_1-\vec\rho)^2(\x_2-\vec\rho)^2}\right)
\;.
\end{equation}
It can be easily seen that the contribution of the virtual part is
$-\left(g_0(\x_1,\x_1)+g_0(\x_2,\x_2)\right)/2$, and taking into
account that
\[
\int \frac{d^2\rho}{(\x_1-\vec\rho)^2(\x_2-\vec\rho)^2}
\left[(\x_1-\x_2)^2 \ln\left(\frac{(\x_1-\vec\rho)^2(\x_2-\vec\rho)^2}
{(\x_1-\x_2)^4}\right)\right.
\]
\begin{equation}
+\left.\biggl((\x_1-\vec\rho)^2-(\x_2-\vec\rho)^2\biggr)
\ln\left(\frac{(\x_1-\vec\rho)^2}{(\x_2-\vec\rho)^2}\right)\right]=0\;,
\end{equation}
we can put $g_0(\x_1,\x_2; \rho)=-f_b(\x_1,\x_2; \rho)$, with $f_b$ given in
(\ref{limit of function f b 1 2}). So, we reached the result of the
previous Section for $D=4$ in a much shorter way.

\section{Conclusion}

The coordinate representation of the BFKL kernel is extremely
interesting, because it gives the possibility to understand its
conformal properties and the relation between the BFKL and the
color dipole approaches. We performed the transformation to the
coordinate representation of the quark contribution to the BFKL
kernel in the next-to-leading order at large $N_c$ from the
momentum representation where it was calculated before. Taking
into account the freedom in the definition of the kernel, we found
agreement with the result obtained recently in Ref.~\cite{Balitsky:2006wa} by
direct calculation of the quark
contribution to the dipole kernel in the coordinate
representation. The agreement is reached after some
transformations of the original BFKL kernel which do not change the scattering
amplitudes of color singlet objects. We have to add that in
Ref.~\cite{Balitsky:2006wa} also terms suppressed by $N_c^2$ are
calculated. These terms have a very complicated form in the momentum
representation~\cite{{FFP99}}. On the contrary, in the coordinate
representation they look quite simple. It would be very
interesting to understand the reason for that.

As for the conformal properties of the NLO BFKL kernel, we have not found any
representation in which the conformal invariance is violated in the NLA only
by the renormalization.

When this article was in the stage of completion, a paper by
Kovchegov and Weigart appeared~\cite{Kovchegov:2006wf}. Our results, obtained
by a quite different approach, agree also with those of that paper.

\vspace{1.0cm} \noindent
{\Large \bf Acknowledgment} \vspace{0.5cm}

V.S.F. thanks the Dipartimento di Fisica dell'Universit\`a della
Calabria and the Istituto Nazionale di Fisica Nucleare, Gruppo
Collegato di Cosenza, for the warm hospitality while part of this
work was done and for the financial support.

\end{document}